# Novel Mechanocaloric Materials for Solid-State Cooling Applications

Claudio Cazorla

*School of Materials Science and Engineering, UNSW Sydney, NSW 2052, Australia*

## *Abstract*

Current refrigeration technologies based on compression cycles of greenhouse gases are environmentally threatening and cannot be scaled down to on-chip dimensions. Solid-state cooling is an environmentally friendly and highly scalable technology that may solve most of the problems associated with current refrigerant methods. Solid-state cooling consists of applying external fields (magnetic, electric, and mechanical) on caloric materials, which react thermally as a result of induced phase transformations. From an energy efficiency point of view, mechanocaloric compounds, in which the phase transitions of interest are driven by mechanical stresses, probably represent the most encouraging type of caloric materials. Conventional mechanocaloric materials like shape-memory alloys already display good cooling performances however in most cases they also present critical mechanical fatigue and hysteresis problems that limit their applicability. Finding new mechanocaloric materials and mechanisms able to overcome those problems while simultaneously rendering large temperature shifts, is necessary to further advance the field of solid-state cooling. In this article, we review novel families of mechanocaloric materials that in recent years have been shown to be specially promising in the aspects that conventional mechanocaloric materials are not, and which exhibit unconventional but significant caloric effects. We put an emphasis on elastocaloric materials, in which the targeted cooling spans are obtained through uniaxial stresses, since from an applied perspective these appear to be the most accomplished. Two different types of mechanocaloric materials emerge as particularly hopeful from our analysis, (1) compounds that exhibit field-induced order-disorder phase transitions involving either ions or molecules (polymers, fast-ion conductors, and plastic crystals), and (2) multiferroics in which the structural parameters are strongly coupled with polar and/or magnetic degrees of freedom (magnetic alloys and oxide perovskites).

## 1. INTRODUCTION

Conventional cooling technologies are based on compression of greenhouse gases that are environmentally threatening (e.g., hydrofluorocarbons). One kilogram of a typical refrigerant gas has the same greenhouse impact than two tonnes of carbon dioxide, which is the equivalent of running a car uninterruptedly for six months [1]. In addition, current cooling technologies present two other critical drawbacks: i) the energy efficiency of refrigeration cycles are low (<60%), and ii) their scalability to small dimensions is very limited. The ~1.4 billion domestic refrigeration units in the world account for about 15% of the total domestic energy consumption, or equivalently, annual $CO_2$ emissions of 450 million tons [2]; hence the potential positive impact in world's sustainability caused by even modest improvements in energy efficiency is enormous. Meanwhile, for microchips to perform optimally the heat generated by electric currents needs to be removed; efficient micro-sized coolers operating near room temperature, therefore, are pressingly needed for successful engineering of faster and more compact electronic devices.

Solid-state cooling is an environmentally friendly, highly energy-efficient, and highly scalable technology that may solve most of the problems associated with current refrigerant methods. Solid-state cooling relies on caloric materials and the application of external fields. In caloric materials reversible temperature shifts are achieved through the application/removal of electric, magnetic, or mechanical fields that render electrocaloric, magnetocaloric, or mechanocaloric effects, respectively. These caloric effects are originated by field-induced phase transitions that involve large changes in entropy (typically, of the order of 10-100 J $kg^{-1}$ $K^{-1}$). Solid-state cooling energy efficiencies of ~75% have been demonstrated and further improvements appear to be within reach [3]. Certainly, the fast response to external stimuli and compactness of caloric materials raise high hopes to surpass the performance, environmental compliance, and portability of current gas-based refrigeration technologies. Caloric materials have actually become a hot topic of research in recent years as it is shown by the rapidly increasing number of related works published in the last decade (Figure 1).

Elastocaloric (EC) and barocaloric (BC) materials lie in the mechanocaloric (MC) category, for which the driving mechanical fields are uniaxial stress and hydrostatic pressure, respectively (e.g., shape-memory alloys). Magnetocaloric compounds (e.g., $Gd_5Si_2Ge_2$) are the most well studied of all caloric materials (Figure 1), and currently they are used in research laboratories for reaching ultralow temperatures (that is, close to the liquefaction points of nitrogen and helium) [4]. The main drawback of magnetocaloric materials is that they require large magnetic fields to drive normal refrigeration spans and usually contain rare-earth elements that are scarce in nature. Meanwhile, electrocaloric materials (e.g., $PbZr_{0.95}Ti_{0.05}O_3$) are well suited for portable cooling applications owing to their high energy density and natural integration in circuitry [5-7]. However, electrocaloric effects usually are too modest, occur at temperatures different from ambient, and require of complex materials synthesis processes in order to avoid impeding leakage currents (even for relatively small electric fields of ~$10^2$ kV/cm) [6].

EC materials, on the other hand, present easy and scalable synthesis along with good availability of the constitutive elements, which leads to affordable production costs. Likewise, the large latent heat accompanying the first-order martensitic phase transition in archetypal EC compounds renders excellent cooling performances [8]. Nevertheless, conventional EC materials also present some drawbacks, mainly associated with their structural stability, mechanical fatigue, and phase-transition hysteresis, which lead to durability and irreversibility issues that may limit severely their cooling performance and applicability [9]. Finding new MC materials with improved mechanical and phase switching properties is highly desirable for the deployment of sustainable, environmentally friendly, and highly scalable solid-state cooling applications.

In this review article, we survey a large number of MC materials ranging from archetypal shape-memory alloys to less renowned EC compounds like polymers, polar oxides, organometal halide perovskites, molecular crystals, multiferroics, and fast-ion conductors. Excellent reviews on elastocaloric and barocaloric materials already exist in the literature [10-13] however the present article, in contrast to those previous ones, focuses on non-conventional MC materials that in recent years have been shown to be particularly promising in the aspects that conventional MC materials are not and which exhibit atypical but sizable MC effects. Our main emphasis is on EC materials since from a technological point of view these are probably the most encouraging, however we also review novel families of barocaloric materials that exhibit remarkable cooling performances and some distinctive field-induced phase transitions. Previous to our EC compounds survey, we present a brief and general description of MC materials, the thermodynamics behind MC effects, and the main techniques that are used to measure and predict MC phenomena. The article ends up with a summary of our most relevant conclusions and outlook.

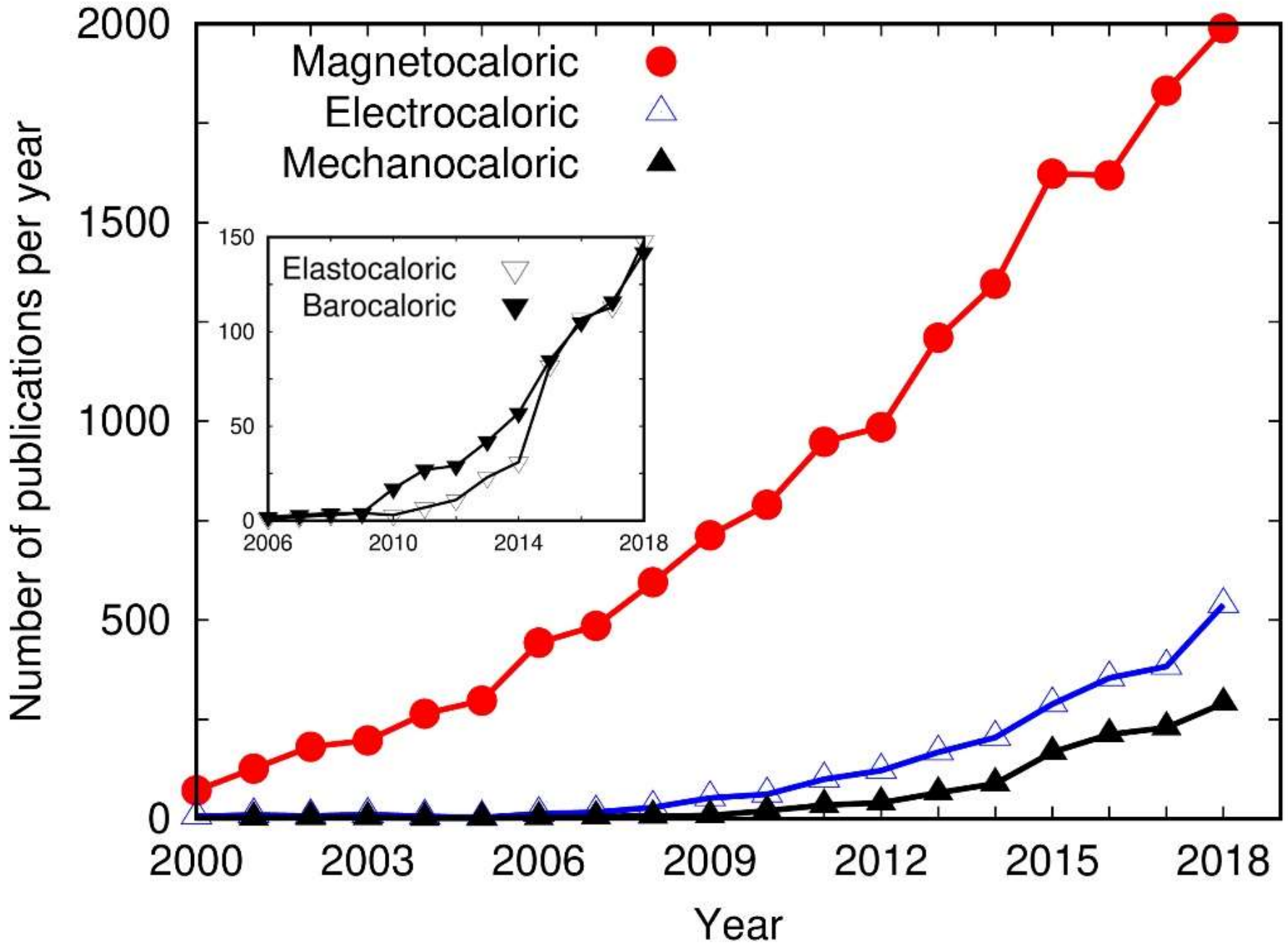

***Figure 1**: Number of articles published on caloric materials since year 2000 (according to Scopus based searches). The tendency in the three general categories of caloric materials is clearly ascendant, especially since year 2010. The number of publications on magnetocaloric materials is overwhelmingly larger, probably due to the fact that these were the first type of compounds for which giant caloric effects were reported [14]. Inset: the number of publications on elastocaloric and barocaloric materials have been very similar in the last years.*

# 2. SOLID-STATE COOLING

## 2.1. EC materials: An overview

Elastocaloric materials undergo reversible changes in temperature when uniaxial mechanical stresses are applied on them under adiabatic conditions. The archetypal class of elastocaloric (EC) effects are those associated to the superelasticity of shape memory alloys (SMA, e.g., non-magnetic Ni-Ti and Cu-based alloys). SMA present a high-temperature high-symmetry phase, known as austenite, and a low-temperature low-symmetry phase, known as martensite, and it is possible to move from one phase to another by means of shear strains. Specifically, a SMA in the austenite phase releases heat into the surroundings when is stressed uniaxially and the martensite phase is stabilized; conversely, the SMA absorbs heat from the surroundings when returning back to the austenite phase upon release of the applied uniaxial stress. Thermal effects associated to the strain-driven austenite-martensite phase transition have been known for about 40 years [15,16], however it has not been until recently that EC effects have been recognized as a potential strategy for room-temperature cooling [17].

Ideal EC materials should render large adiabatic temperature spans, $\Delta T$ (~10K), under the application of small or moderate uniaxial stresses (typically $10^2$-$10^3$ MPa) at (or near) room temperature. A parameter that quantifies the cooling performance of EC materials is the cooling/heat pumping cycle efficiency (*COP*), which is defined as:

$$COP = \frac{|Q|}{W} = \frac{m \cdot c \cdot |\Delta T|}{\frac{1}{\rho} \oint \sigma \cdot d\varepsilon} \qquad (1)$$

where $|Q|$ and $W$ represent the amount of exchanged heat and mechanical work done on the material, respectively, $m$ the molar mass, $c$ the molar heat capacity, $\rho$ the density, $\sigma$ and $\varepsilon$ and the applied uniaxial stress and resulting uniaxial strain. Typical *COP* values of efficient EC materials lie within the range of 2-8 [18]. Another useful refrigerant performance descriptor is the so-called refrigerant capacity (*RC*), which is defined as:

$$RC = \int_{T_0}^{T_1} \Delta S \cdot dT \ , \qquad (2)$$

where $T_0$ and $T_1$ represent the lower and upper temperatures delimiting the interval in which the caloric effects of interest are observed. Likewise, one can define the normalised refrigerant capacity (*NRC*) which simply is the RC divided by the involved change in external field, $\Delta X$ ($\Delta\sigma$ in the case of EC effects):

$$NRC = \frac{RC}{\Delta X} \ . \qquad (3)$$

Typical *RC* and *NRC* values of good MC materials are of the order of 1 kJ $kg^{-1}$ (5 J $cm^{-3}$) and 10 kJ $kg^{-1}$ $GPa^{-1}$, respectively [10,19].

A number of prototypes based on EC materials already exist, which are mostly based on Ni-Ti SMA (for a detailed description of such prototypes see [20]). Typical values of device *COP* (which normally are significantly lower than the corresponding material *COP* - see Eq.(1) - due to irreversible heat transfer losses that occur in practice) and pumping power are around unity and ~10 W, respectively. Those prototypes can be classified into two broad categories depending on the method used for transferring heat from the cool to the heat source. The first class of prototypes refers to devices in which the EC compound remains stationary and a fluid is circulated to facilitate heat transfer [21,22]. In the second category, the EC element is mobile and heat transfer is realized by direct contact with the cool and heat sources [23]. Active elastocaloric refrigerators (i.e., involving heat exchangers, EC plates, a heat-transfer fluid, and an actuator to load/unload the material) have been already designed [24], and a regenerative EC device has been recently shown to exceed in performance previous refrigerant devices based on the magnetocaloric effect [25]. Although there still are few technical aspects that need to be improved in those prototypes (e.g., cyclability, materials fatigue, and heat transfer), they very well illustrate the great potential of elastocaloric materials in the context of solid-state cooling.

## 2.2. The elastocaloric effect

When uniaxially stressed, EC materials undergo a change in temperature under adiabatic conditions, $\Delta T$, or equivalently a change in entropy under isothermal conditions, $\Delta S$. The entropy of a system subject to an external uniaxial stress, $\sigma$, depends on $\sigma$ and the temperature, $T$, hence an infinitesimal change in entropy can be expressed as:

$$dS = \frac{\partial S}{\partial \sigma} d\sigma + \frac{\partial S}{\partial T} dT \,. \tag{4}$$

The partial derivative of the system entropy with respect to temperature is related to the heat capacity, $C$ (at fixed stress), as:

$$\frac{\partial S}{\partial T} = \frac{C}{T} \,. \tag{5}$$

Likewise, the Maxwell relation:

$$\frac{\partial S}{\partial \sigma} = V_0 \frac{\partial \varepsilon}{\partial T} \,, \tag{6}$$

allows to express the stress derivative of the system entropy as a function of its $T$-induced elongation, $\varepsilon$, at fixed uniaxial stress ($V_0$ corresponds to the equilibrium system volume). By considering the thermodynamic relations (5) and (6), one can rewrite $dS$ in a physically more insightful manner like:

$$dS = V_0 \frac{\partial \varepsilon}{\partial T} d\sigma + \frac{C}{T} dT \,. \tag{7}$$

From Eq. (7), it is straightforward to deduce the analytical expression of the entropy change that an EC material experiences upon application of uniaxial stresses at fixed temperature (that is, $dT = 0$), namely:

$$\Delta S = V_0 \int_0^{\sigma} \frac{\partial \varepsilon}{\partial T} d\sigma' \,, \tag{8}$$

as well as the expression of the corresponding temperature change occurring under adiabatic conditions (that is, $dS = 0$), which is:

$$\Delta T = -V_0 \int_0^{\sigma} \frac{T}{C} \frac{\partial \varepsilon}{\partial T} d\sigma' \,. \tag{9}$$

When temperature changes are small and the heat capacity of the system does not change appreciably with the applied uniaxial stress, the adiabatic temperature change of an EC material can be reasonably approximated by:

$$\Delta T \approx -\frac{T}{C}\,\Delta S\,. \qquad (10)$$

The equation above shows that a negative isothermal entropy shift leads to a positive adiabatic temperature shift (heating, direct EC effect), and conversely, a positive $\Delta S$ leads to a negative $\Delta T$ (cooling, inverse EC effect). EC effects associated to SMA normally are direct during stress loading. When the phase transition rendering EC effects is of first-order type, one can use the Clausius-Clapeyron method to express the accompanying isothermal entropy change as [26]:

$$\Delta S = -V_0\,\Delta\varepsilon\,\frac{\partial\sigma}{\partial T}\,, \qquad (11)$$

where, $\Delta\varepsilon$ represents the change in length that the material experiences during the transformation along the direction of the applied stress. We note that the Clausius-Clapeyron method disregards possible caloric effects arising in each of the involved phases at thermodynamic conditions other than the transition point (which usually are small).

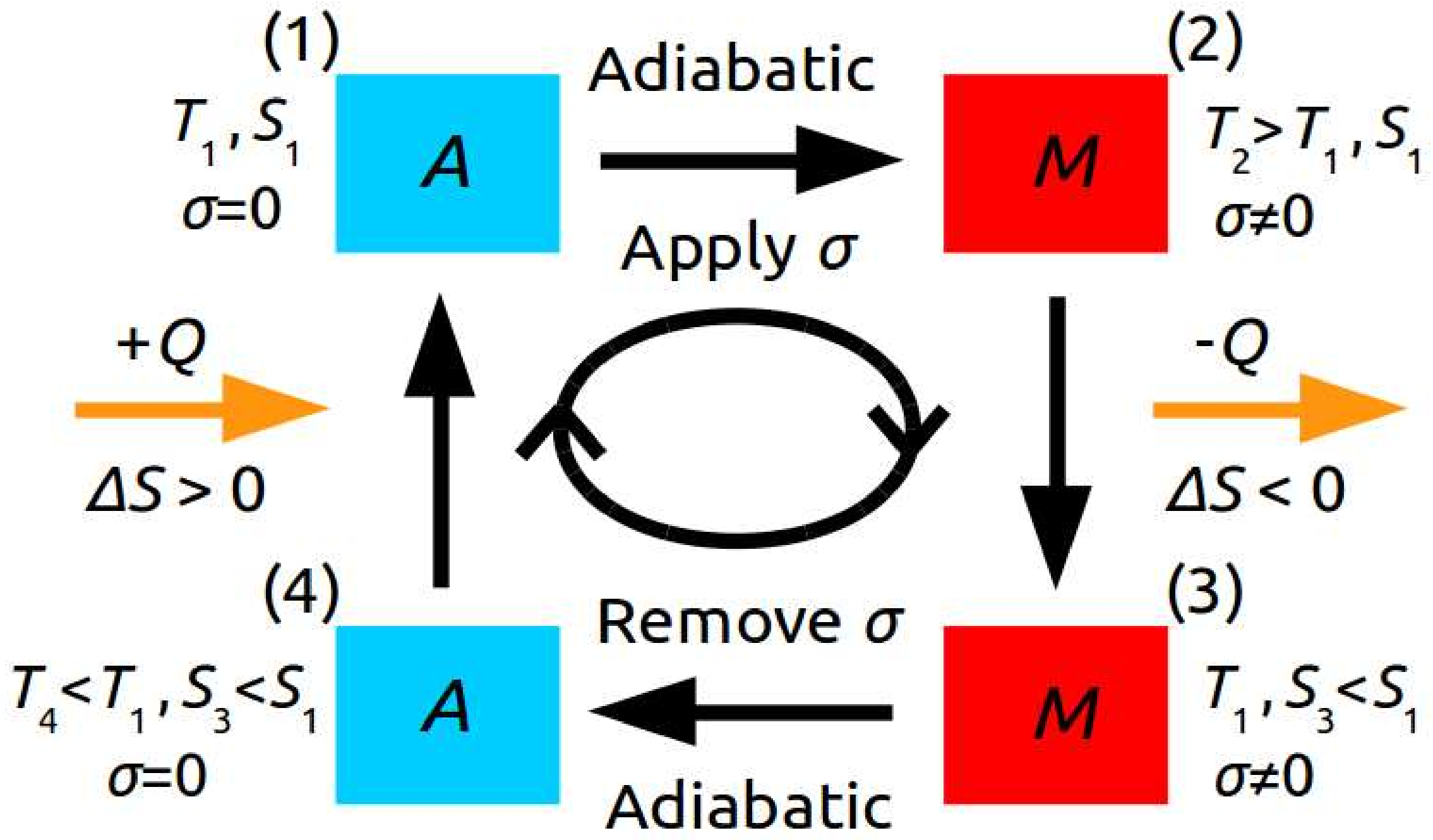

***Figure 2**: Sketch of the typical four-steps refrigerant cycle based on EC materials (see main text). "A" stands for the initial low-symmetry low-T phase of the system in the absence of any stress (e.g., austenite phase in SMA), and "M" for the corresponding strain-induced high-symmetry high-T phase (e.g., martensite phase in SMA).*

A simple four-steps cooling cycle based on EC materials (and in general on any caloric compound) can be envisaged consisting of two adiabatic and two non-isothermal processes (Figure 2). Starting from the low-symmetry low-$T$ phase at temperature $T_1$, a uniaxial stress is applied adiabatically on the sample that stabilizes the corresponding high-symmetry high-$T$ phase. Consequently, the temperature of the EC element increases up to $T_2 = T_1 + \Delta T$, where the size and sign of $\Delta T$ depends on the material and applied stress [see Eq.(9)]. The EC material, which still remains elastically strained, then is put into contact with a thermal sink so that heat is ejected from the system. Consequently, the temperature of the EC element returns back to $T_1$ and its entropy becomes smaller than it was at the beginning of the cycle. Subsequently, the uniaxial stress is adiabatically removed and the temperature of the EC element falls down to $T_4 = T_1 - \Delta T$. Finally, the EC component is put into contact with the hot source that needs to be cooled down so that heat is absorbed by the system and the initial values of the temperature and entropy are restored.

## 2.3. Experimental and theoretical measurement techniques

In practice, the temperature and entropy increments associated to EC effects are measured with direct or indirect methods. Direct methods refer to *in situ* temperature measurements on the samples during the stress loading/unloading cycles. Indirect methods, on the other hand, involve strain measurements under varying uniaxial stress and temperature conditions that allow to estimate $\Delta T$ and $\Delta S$ through the Maxwell relations (e.g., Eq.(8)). When the phase transition rendering EC effects is of first-order type, one may use the Clausius-Clapeyron method (e.g., Eq.(11)); this is a particular case of indirect methods that involve the determination of transition temperatures during cooling and warming cycles. Direct methods (e.g., thermocouples mounted on the samples and thermography IR cameras) are more straightforward than indirect methods however they are also more prone to be affected by systematic errors (actually, caloric data obtained with direct and indirect methods normally differ [27]). For instance, since the EC material needs to be connected to an actuator-like element for stress loading and unloading, in practice there will always be heat leakages to the ambient that will lead to irreversible processes (i.e., non-adiabaticity) and consequently to $\Delta S$ underestimation. There is a third category of measurement techniques named quasi-direct methods that involves differential scanning calorimetry to measure directly heat transfers (which can then be converted straightforwardly into entropies). This last type of techniques is particularly well suited for performing barocaloric measurements but entails some technical challenges when it comes to measuring elastocaloric effects.

A different way of estimating EC effects is with theoretical methods. Theoretical methods are inexpensive from an economical cost point of view, can be very accurate, and have predictive power, hence they may serve as a guide for the experiments. By theoretical methods we refer to first-principles methods (e.g., density functional theory [28,29]), atomistic simulation techniques (e.g., classical molecular dynamics and Monte Carlo methods - see, for instance, [30] -), and phenomenological and effective Hamiltonian models (see, for instance, [24] and [31]). A particularly promising computational technique that allows to estimate phase-transition entropies and heat capacities in crystals is the quasi-harmonic (QH) approach [32,33]. In the QH approach, one computes first the vibrational frequencies of a material, $\omega_{\mathbf{q}s}$, (with the "small displacement" method, for instance [34,35]) and subsequently obtains the corresponding vibrational entropy, $S_{vib}$, and heat capacity, $C_V$, through the exact analytical expressions:

$$S_{vib}(T) = -\frac{1}{N_{\mathbf{q}}} k_B \sum_{\mathbf{q},s} \ln\left[2\sinh\left(\frac{\hbar\omega_{\mathbf{q}s}}{2k_BT}\right)\right], \tag{12}$$

and

$$C_V(T) = \frac{1}{N_{\boldsymbol{q}}} \sum_{\boldsymbol{q},s} \frac{\left(\hbar\omega_{\boldsymbol{q}s}\right)^2}{k_B T^2} \cdot \frac{e^{\frac{\hbar\omega_{qs}}{k_B T}}}{\left(e^{\frac{\hbar\omega_{qs}}{k_B T}} - 1\right)^2}\ , \qquad (13)$$

where $N_{\boldsymbol{q}}$ is the total number of wave vectors used for integration within the Brillouin zone and the summation runs over all wave vectors $\boldsymbol{q}$ and phonon branches $s$. The main disadvantage of the QH approach is that strong anharmonic effects are completely disregarded, hence one needs to check explicitly in each case whether this approximation compromises or not the reliability of the obtained theoretical results [36-38]. In addition, possible sources of entropy other than lattice vibrations need to be considered separately (e.g., stemming from the magnetic, electronic, and/or electric degrees of freedom) [7,39].

Other less conventional types of computational approaches based on high-throughput searches, which are becoming increasingly more popular within the community of materials scientists, have been employed also recently for the design of new caloric materials [40,41]. A detailed description of the mentioned theoretical methods escapes from the scope of the present review but we will refer to the relevant works when necessary.

# 3. ARCHETYPAL EC MATERIALS: SHAPE-MEMORY ALLOYS

Shape-memory alloys (SMA) are archetypal ferroelastic materials in which the lattice distortion is dominated by shear strain. Generally, SMA possess very good ductility: they can be strained by about 10% and more. These superb elastic properties along with the large latent heat associated with the austenite to martensite phase transition ($|\Delta S| \sim 10$ J kg$^{-1}$ K$^{-1}$) convert SMA into excellent EC materials. SMA can be divided into two broad groups, namely, magnetic (e.g., Ni-Mn-Sb-Co and Ni-Fe-Ga alloys) and non-magnetic (e.g., Ni-Ti based, Cu-based, and Fe-based alloys). The most widely investigated class of EC materials are near-equiatomic Ni-Ti alloys (Table 1). The high-temperature phase of Ni-Ti alloys, usually denoted as austenite, presents cubic symmetry (Figure 3) and is easily deformable by shearing the {110} planes along the <1-10> direction. The phonon branch associated to such a distortion (labelled as $TA_2$) and the related elastic constant ($C'$) are both *soft*, which leads to high vibrational entropy and the stabilization of the austenite phase at high temperatures [42]. The phase transition to the martensite phase is diffusionless and of first-order type. The low-temperature martensite phase in Ni-Ti alloys is monoclinic (space group *P21/m*, Figure 3), however other intermediate structural phases may be observed in the experiments depending on the heat treatment and chemical composition of the material samples.

In Table 1, we summarize the properties of some SMA-based compounds in which giant EC effects (i.e., $|\Delta S| \sim 10$ J kg$^{-1}$ K$^{-1}$ and $|\Delta T| \sim 10$ K) have been observed. For instance, an adiabatic temperature change of 25 K (21-17 K) under application (removal) of a ~500 MPa stress tension has been measured in Ni-Ti wires [43,44]. Likewise, a $|\Delta T|$ of 17 K (16 K) under stress loading (unloading) has been observed in a Ni-Ti thin film [45].

SMA alloys present two main drawbacks in the context of EC cooling [9,10]. One is related to the unavoidable hysteresis effects associated with the first-order nature of the martensitic phase transition. Hysteresis originates from the fact that the transition path occurs under non-equilibrium conditions, hence irreversible processes take place within the material. From a practical point of view, irreversibility may pose severe limitations to the overall cooling performance attained during successive field-induced cycles. In particular, the irreversibility of any caloric effect depends on two parameters: (i) the temperature width of the hysteresis and (ii) the temperature-shift of the entire hysteresis loop due to application/removal of the external field. Certainly, the narrower the hysteresis is the smaller the irreversibility of the process. On the other hand, if the width of the hysteresis is significant, a large temperature shift of the characteristic transition temperature as induced by the application of the external field may lead to a considerable reduction in irreversibility. Recently, it has been shown in magnetic SMA that a possible way of improving the reversibility of cooling effects, or even of exploiting hysteresis, may consist in combining different types of external stimulus (e. g., magnetic and mechanical) [46]; we will comment again in this case later on when reviewing multiferroics.

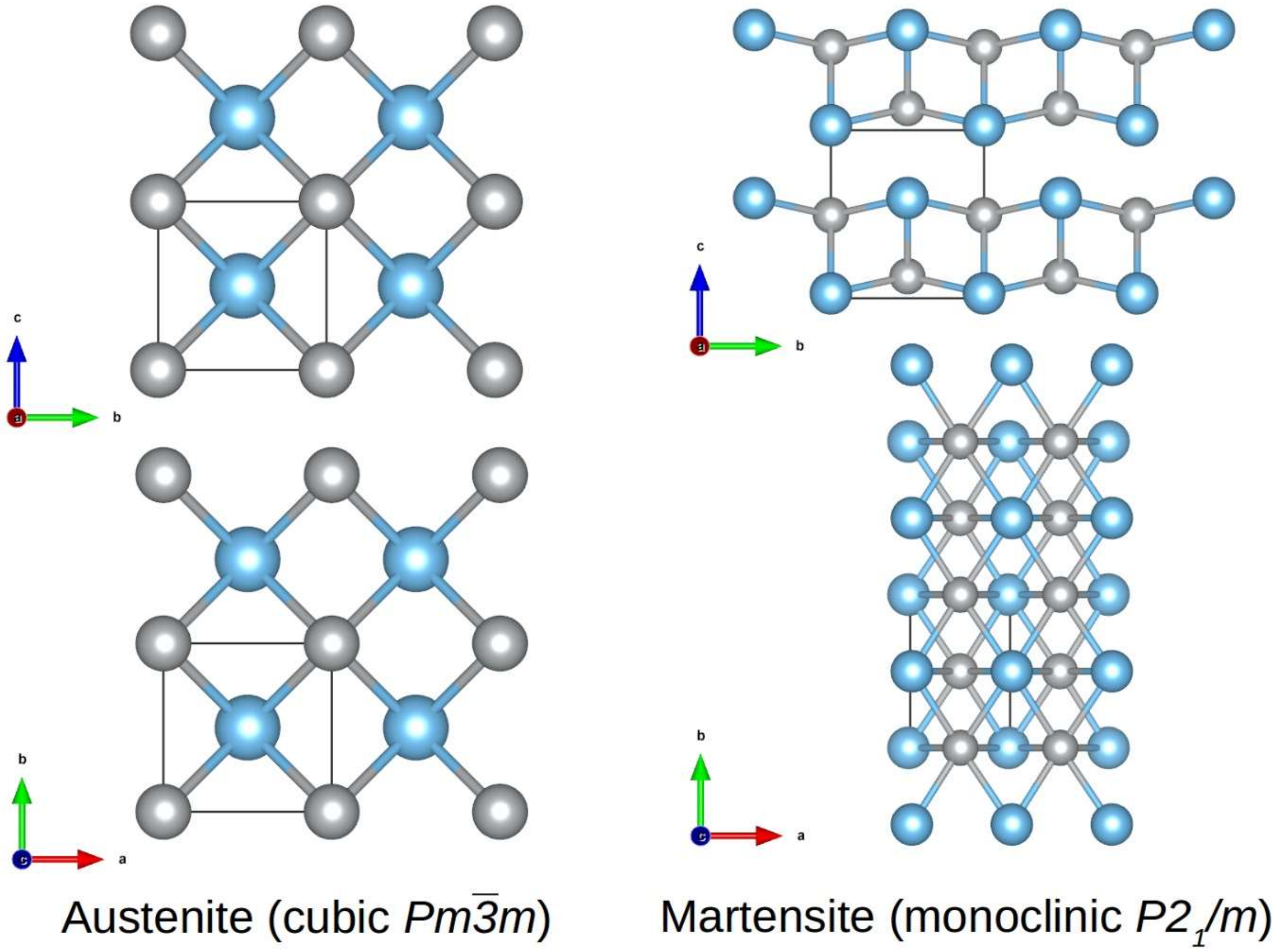

***Figure 3**: Ball-stick representation of the usual high-temperature high-symmetry austenite and low-temperature low-symmetry martensite phase in Ni-Ti based alloys. Ni and Ti atoms are represented with grey and blue spheres, respectively.*

Another non-negligible drawback of SMA alloys is that related to their mechanical stability and fatigue. This family of materials display two different types of fatigue, (i) structural fatigue, as originated by conventional cracks formation and propagation eventually leading to fracture, and (ii) functional fatigue, linked to the loss of superelastic properties over increasing number of cycles. These issues typically limit the durability of SMA alloys to just few thousands of cycles, which is insufficient for practical applications (e. g., heat-pumping devices).

Aimed at improving the structural stability and mechanical fatigue properties of Ni-Ti alloys, doping strategies based on Cu, Co, Fe, and V have been explored. In Ni-Fe-Ga and Ni-Co-Al alloys, for instance, adiabatic temperature spans of 8.4 and 3.1 K have been measured directly, while values about a factor of 2 larger have been estimated indirectly [47]. An adiabatic temperature change of 10 K has been observed in Ni-Ti-Cu-Co thin films, which remarkably can endure several millions of loading-unloading strain cycles of $|\varepsilon| \sim 2.5\%$ [48,49]. Among Cu-based SMA, Cu-Zn-Al has received special attention; a $|\Delta T|$ of 15 K has been measured indirectly for a small uniaxial stress of 28.5 MPa [17,50,51]. A negative adiabatic temperature shift of 6 K (i. e., inverse EC effect) has been also observed in Cu-Zn-Al alloys at temperatures below ambient. Among Fe-based SMA, Fe-Rh and Fe-Pd alloys have been found to display large and giant EC effects; adiabatic temperature changes of 5-8 K have been reported in Fe-Rh near room temperature [52], and of 2 K in Fe-Pd under moderate uniaxial stresses (~500 MPa) [53].

| **Compound** | ***C*** [J kg$^{-1}$ K$^{-1}$] | $T_t$ [K] | $|\varepsilon|$ [%] | $|\Delta S_t|$ [J kg$^{-1}$ K$^{-1}$] | $|\Delta T_t|$ [K] | $|\Delta T_{(d)}|$ [K] | $|\Delta\sigma|$ [MPa] | **Reference** |
|---|---|---|---|---|---|---|---|---|
| $Ni_{50}Ti_{50}$ (w) | 550 | 300 | 8 | 40 | 22 | 25 | 500 | [43] |
| $Ni_{50}Ti_{50}$ (f) | 420 | 260 | 3.5 | 77 | 48 | 16 | 500 | [45] |
| $Ni_{45}Ti_{47}Cu_5V_3$ | 500 | 250 | 5 | 40 | 20 | 8 | 500 | [54] |
| $Ni_{31}Ti_{55}Cu_{12}Co_2$ | 420 | 280 | 2 | 30 | 20 | 10 | 200 | [49] |
| $Ni_{54}Fe_{19}Ga_{27}$ | 470 | 280 | 3 | 11 | 6 | 4 | 170 | [55] |
| $Ni_{54}Fe_{19}Ga_{27}Co_4$ | 450 | 300 | 5 | 11 | 7 | 10 | 300 | [56] |
| $Ni_{33}Co_{40}Al_{29}$ | 480 | 290 | 10 | 44 | 26 | 3 | 50 | [47] |
| $Ni_{43}Mn_{40}Sn_{10}Cu_7$ | 450 | 320 | 1 | 54 | 38 | - | 10 | [57] |
| $Ni_{46}Mn_{38}Sb_{12}Co_4$ | 400 | 300 | 3 | 34 | 25 | - | 100 | [58] |
| $Ni_{45}Mn_{44}Sn_{11}$ | 580 | 270 | 2 | 32 | 15 | 6 | 260 | [59] |
| $Ni_{45}Mn_{36}In_{14}Co_5$ | 460 | 250 | 3 | 18 | 10 | 4 | 150 | [60] |
| $Ni_{48}Mn_{35}In_{17}$ | 400 | 300 | 2 | 40 | 30 | 4 | 250 | [61] |
| $Cu_{68}Zn_{16}Al_{16}$ | 430 | 230 | 8 | 21 | 22 | 6 | 120 | [17,50] |
| $Cu_{83}Al_{14}Ni_3$ | 410 | 250 | 8 | 21 | 13 | 16 | 150 | [15] |
| $Fe_{49}Rh_{51}$ | 470 | 310 | 0.3 | 13 | 8 | 5 | 500 | [52] |
| $Fe_{69}Pd_{31}$ | 400 | 250 | 2 | 1 | 0.6 | 3 | 200 | [53] |

***Table 1**: SMA in which giant EC effects have been observed. $T_t$ represents operating temperature, C heat capacity of the compound, ε uniaxial strain, Δσ applied uniaxial stress, $\Delta S_t$ entropy change associated to the phase transition, $\Delta T_t$ adiabatic temperature change deduced from the entropy change associated to the phase transition, and $\Delta T_{(d)}$ adiabatic temperature change that is measured directly. '(w)' stands for wire and '(f)' for thin film.*

Likewise, large EC effects have been observed in magnetic SMA such as Ni-Fe-Ga [55,56] and Ni-Mn-Sb-Co [57,58] alloys near room temperature. For instance, an adiabatic temperature change of 10 K has been measured directly in $Ni_{54}Fe_{19}Ga_{27}Co_4$ at $T$ = 300 K and $|\Delta\sigma|$ = 300 MPa [56]. Nevertheless, magnetic SMA alloys, in contrast to non-magnetic SMA, are brittle and consequently cannot withstand large uniaxial deformations.

As regards possible cooling applications, the temperature span over which large $\Delta S$ and $\Delta T$ are observed should be as large as possible. The refrigerant cooling performance (*RCP*), defined as $RCP = |\Delta S_{\mathrm{peak}}|$ x |FWHM of $\Delta S(T)$| where "FWHM" stands for full width at half maximum (hence it has units of temperature, see Eq. (2)), is a good indicator of such a quality. In a recent study of a Cu-Zn-Al polycrystal [51] an outstanding refrigerant cooling performance of 2300 J $kg^{-1}$ has been measured, which in fact demonstrates the great promise of SMA for realizing practical mechano-cooling devices.

# 4. NON-CONVENTIONAL EC MATERIALS

In the last years, families of materials other than SMA have been reported to exhibit also large or giant EC effects (Table 2). Examples include organic-inorganic polymers [62-64], polar materials [65-69], fast-ion conductors [70-72], hybrid perovskites [73], and multiferroics [46]. Although SMA may outperform such non-conventional EC materials in some facets, the latter present huge room for improvement, as they have not been optimized or analysed exhaustively yet, and are particularly promising in the aspects that SMA are not (i.e., huge $|\Delta T|$'s, high mechanical stability, and low hysteresis). In this section we explain and summarize the properties of non-conventional EC compounds (which have not been considered in detail in previous review articles), in an attempt to broaden the scope of mechanocaloric materials and of solid-state cooling in general.

| **Compound** | **Material type** | ***$T_t$*** [K] | **\|*ε*\|** [%] | **\|*$\Delta S_t$*\|** [J kg$^{-1}$ K$^{-1}$] | **\|*$\Delta T_t$*\|** [K] | **\|*$\Delta\sigma$*\|** [MPa] | **Reference** |
|---|---|---|---|---|---|---|---|
| Natural Rubber | POL | 300 | 600 | 80 | 9-12 | 1-2 | [62,75] |
| PVDF | POL | 300 | 12 | 11 | 2 | 15 | [63,64] |
| VNR | POL | 300 | 6 | 87 | 10 | 173 | [77] |
| PDMS | POL | 300 | 6 | 70 | 12 | 173 | [78] |
| BCZTO | FE | 340 | 0.2 | - | 2 | 250 | [67] |
| $BaTiO_3$ | FE | 300 | 3 | 8 | 5 | 6,500 | [65,66] |
| $Ba_{0.5}Sr_{0.5}TiO_3$ | FE | 260 | 1.2 | - | 9 | 1,000 | [31] |
| $PbTiO_3$ | FE/MF | 700 | 1.5 | - | 35 | 1,000 | [79] |
| $PbZrO_3$ | AFE/MF | 950 | 0.8 | - | 25 | 2,000 | [80] |
| $CaF_2$ (f) | FIC | 1,350 | 2.8 | 200 | 163 | 5,000 | [70] |
| $PbF_2$ (f) | FIC | 600 | 0.5 | 60 | 22 | 600 | [70] |
| AgI (f) | FIC | 300 | 10 | 30 | 38 | 1,000 | [71] |
| $LiIO_3$ (f) | FIC | 1,000 | 2 | 16 | 9 | 1,000 | [71] |
| $Li_3N$ | FIC | 300 | 2 | 20 | 2 | 5,800 | [72] |
| $CH_3NH_3PbI_3$ | HP | 300 | 1 | - | 11 | 550 | [73] |
| $Ni_{49.6}Mn_{35.6}In_{14.8}$ | MF | 293 | 3.5 | - | 1.3 | 75 | [46] |
| $FeRh/BaTiO_3$ | MF | 390 | - | 8 | 5.2 | - | [113] |

| | | | | | | | |
|---|---|---|---|---|---|---|---|

***Table 2**: Materials other than SMA in which large EC effects have been observed or predicted. $T_t$ represents operating temperature, ε uniaxial strain, Δσ applied uniaxial stress, $\Delta S_t$ entropy change associated to the phase transition, and $\Delta T_t$ adiabatic temperature change deduced from the entropy change associated to the phase transition. '(f)' stands for thin film, 'PVDF' for polyvinylidene di-fluoride polymers, 'VNR' for vulcanized natural rubber, and 'PDMS' for polydimethylsiloxane. 'POL', 'FE', 'AFE', 'FIC', 'HP', and 'MF' indicate organic-inorganic polymers, polar materials, anti-ferroelectric, fast-ion conductors, hybrid perovskites, and multiferroic, respectively.*

## 4.1. Hybrid organic-inorganic polymers

Natural rubber was the first material in which elastocaloric effects were observed [74]. Recently, the EC properties of this material have been revisited. Natural rubber is an elastomer consisting of polymeric chains randomly oriented. Upon application of uniaxial stress the polymeric chains in natural rubber become ordered along the direction of the strain, which produces a decrease in the configurational entropy of the system (hence the origin of the observed EC effect). Adiabatic temperature changes of 9-12 K have been measured directly in natural rubber near room temperature for huge uniaxial strains of ~ 600% (resulting from minute uniaxial stresses of just 1-2 MPa) [62,75].

Polymers based on polyvinylidene fluoride (PVDF) also conform to a promising family of EC materials. These polymers are ferroelectrics, that is, exhibit spontaneous and switchable electric polarization, and the accompanying polar to non-polar phase transition is very sensitive to external mechanical stresses (hence the origin of the reported EC effects). Giant electrocaloric effects have been reported also recently in these materials [76]. Adiabatic temperature changes of ~2 K have been measured directly in PVDF polymers near room temperature under moderate uniaxial strains of ~10% (equivalent to small uniaxial stresses of ~10 MPa). The reported adiabatic temperature changes in PVDF polymers normally are large but not giant essentially due to the huge heat capacity of such light-weight materials (i.e., $C \sim$ 1,700 $JK^{-1}kg^{-1}$ [76] to be compared with typical SMA values of ~ 500 $JK^{-1}kg^{-1}$, Table 1).

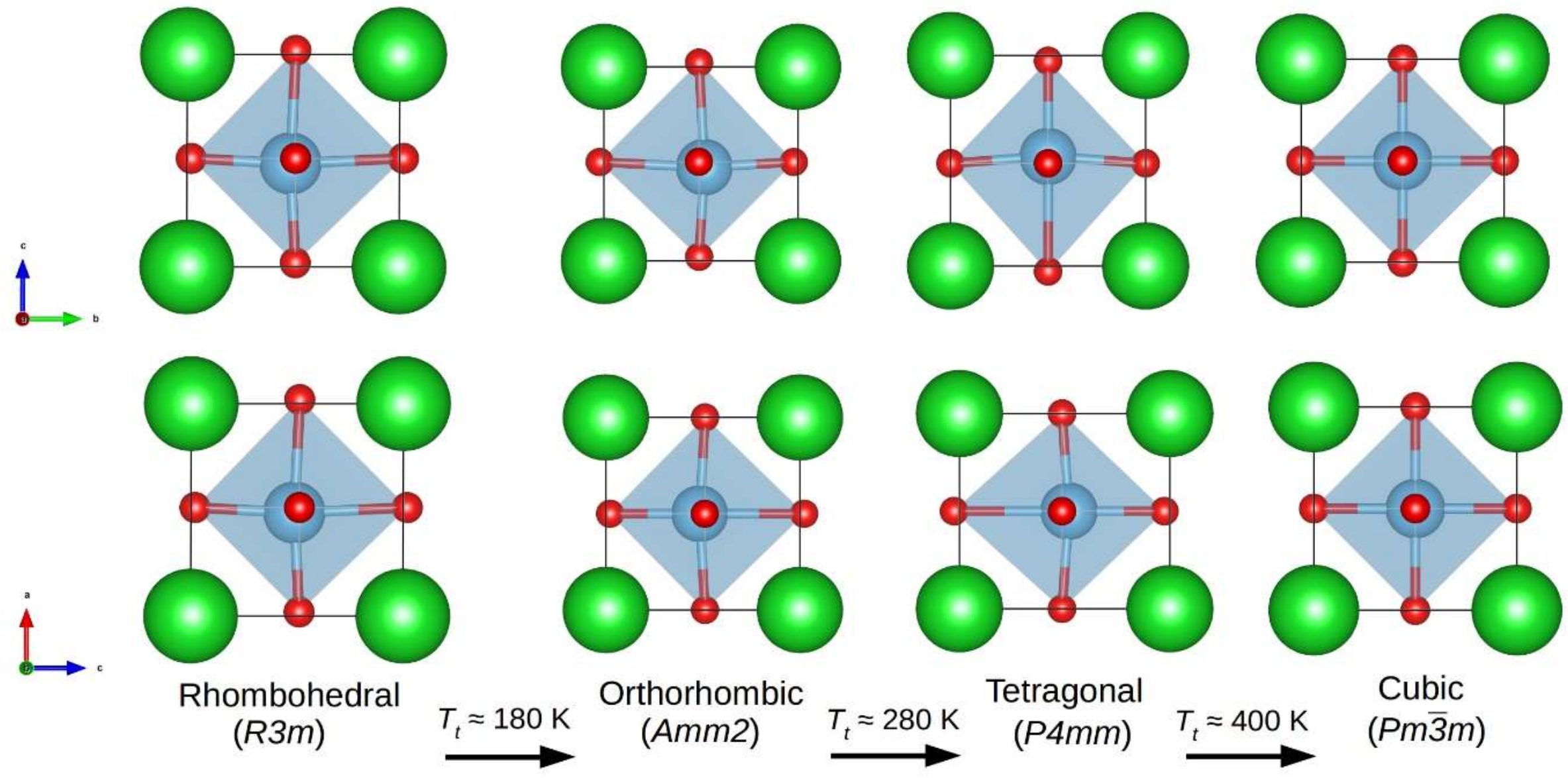

***Figure 4****: Representation of the phase transition sequence occurring in the archetypal ferroelectric compound $BaTiO_3$ under increasing temperature. Ba, Ti, and O atoms are represented with green, blue, and red spheres, respectively. The rhombohedral, orthorhombic, and tetragonal phases are polar whereas the cubic phase is non-polar (i.e., centrosymmetric). The space group of each relevant phase is indicated within parentheses along with the involved transition temperatures.*

More recently, large elastocaloric effects around room temperature have been reported for vulcanized natural rubber (VNR) [77] and polydimethylsiloxane (PDMS) [78]. One advantage of elastomers is that they exhibit giant EC effects around room temperature even in the absence of conventional structural phase transitions. Consequently, the degree of reversibility associated with such EC effects is very high and hysteresis and phase coexistence issues should be practically absent in cooling cycles based on them.

## 4.2. Ferroelectric oxide perovskites

Ferroelectric (FE) oxide perovskites are piezoelectric materials. Piezoelectricity is the quality by which an electrical current can be created in a material through the application of a mechanical stress and *vice versa*. Ferroelectric oxide perovskites, therefore, generally are very sensitive to strains [81]. In addition, FE oxide perovskites may endure large mechanical deformations when finely synthesized as thin films (e.g., a huge and reversible strain of 14% driven either by an electric field or temperature has been experimentally observed in epitaxially strained multiferroic $BiFeO_3$ [82]). Moreover, polar to non-polar phase transitions occurring in FE oxide perovskites typically are soft-mode driven and of second-order (or weak first-order) type; consequently, accompanying hysteresis effects leading to irreversibility may turn out to be relatively small.

The archetypal FE oxide perovskite is $BaTiO_3$ (BTO), which at low temperatures presents a rhombohedral phase characterized by an electric polarization oriented along the pseudo-cubic direction [111] (Figure 4). As the temperature is raised, BTO undergoes a series of abrupt electro-structural phase transformations: first from the rhombohedral (*R*) ground state to an orthorhombic (*O*) phase ($T_t$ = 180 K), with the electric polarization along [110], subsequently to a tetragonal (*T*) phase ($T_t$ = 280 K), with the electric polarization along [001], and finally to a non-polar cubic (*C*) phase ($T_t$ = 400 K) [83] (Figure 4). Such a sequence of *T*-induced phase transformations, or portions of it, is observed in many other functional materials, including perovskite-structured compounds (e.g., $CH_3NH_3PbI_3$ [84]) and ferroelectric relaxors (e.g., $(Na_{0.5}Bi_{0.5})_{1-x}Ba_xTiO_3$ [85]). Interestingly, the *R-O-T-C* sequence of phase transitions can be induced also with mechanical stresses at fixed temperature [28,86]. The significant changes in electrical polarization and structural parameters occurring during such a phase transition sequence lie at the heart of most caloric effects observed in FE oxide perovskites.

Small EC effects have been predicted in $BaTiO_3$ at room temperature, in particular, a $|\Delta T|$ ~ 5 K for huge uniaxial stresses of several thousands of MPa (Table 2) [65,66]. A similar adiabatic temperature change has been measured in a related oxide compound, namely, $(Ba_{0.85}Ca_{0.15})(Zr_{0.1}Ti_{0.9})O_3$, although in this case for a significantly smaller uniaxial stress ($\sigma$ ~ 250 MPa) [67]. Of particular promise are the EC effects predicted in $Ba_{0.5}Sr_{0.5}TiO_3$ near room temperature, i.e., $|\Delta T|$ ~ 10 K, since these are partly originated by the strong coupling between the elastic and polar degrees of freedom in the material, a common trait in many ferroelectric perovskites [31]. Exploitation of the coupling between several ferroic orders that coexist in a same compound (e.g., elastic, electric, and magnetic), usually referred to as multiferroicity, in fact has been suggested as a possible route for achieving giant caloric effects [87]. In this context, a giant adiabatic temperature change of ~ 35 K has been also predicted in the well-known ferroelectric material $PbTiO_3$ (PTO), which has been ascribed to both the elastic and ferroelastic responses of the compound [79]. Unfortunately, in this latter case the transition temperature associated to the EC effects is too high (~700 K, although large room-temperature EC effects have been predicted also in PTO polycrystals [88]) and the material in question contains toxic elements (i.e., Pb). For completion purposes, we report in Table 2 the EC effects that have been estimated theoretically in the anti-ferroelectric compound $PbZrO_3$ [80], which have a similar origin to those disclosed in PTO.

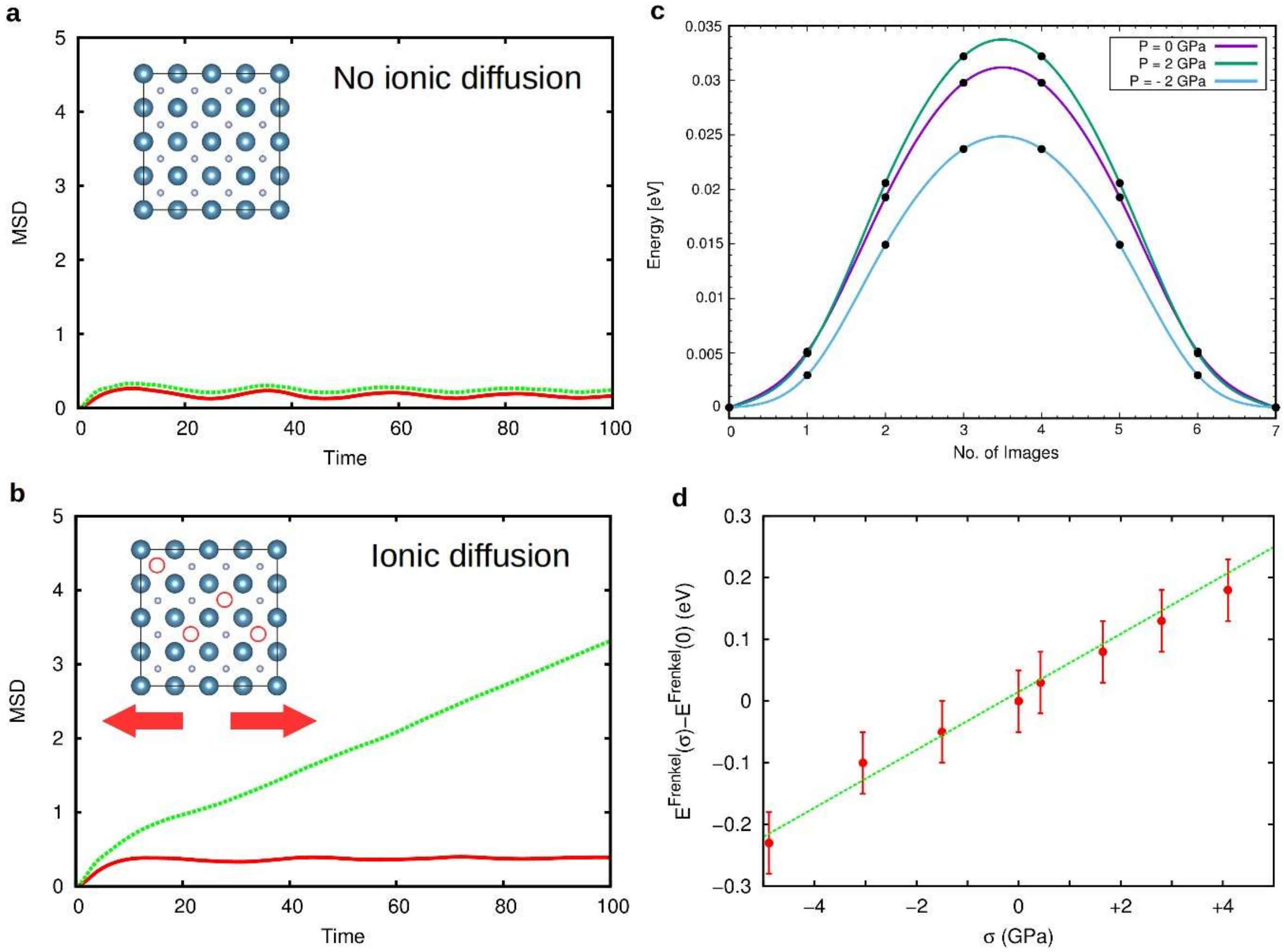

***Figure 5***: *Representation of the ionic-conductivity enhancement effect driven by mechanical stress in fast-ion conductors.* ***a.*** *A FIC is in the normal crystalline state.* ***b.*** *Application of tensile stress stabilizes the superionic state without the need of increasing the temperature. The σ-induced ionic conductivity enhancement effect is originated by the lowering of* ***c.*** *kinetic energy barriers involved in ionic diffusion, and* ***d.*** *formation energy of Frenkel pair defects. Conversely, compressive stresses tend to deplete ionic conductivity.*

## 4.3. Fast-ion conductors

Fast-ion conductors (FIC) are materials that exhibit very high ionic conductivity in the crystal phase at temperatures well below their corresponding fusion points [89]. Examples of FIC, which are also known as superionic conductors, include metal fluorides (e.g, $CaF_2$ and $PbF_2$), metal iodides (e.g., AgI), oxides (doped $CeO_2$), metal chalcogenides (e.g., $Cu_2Se$ and $Ag_2S$), and lithium-based (e.g., $Li_3N$ and $Li_{10}GeP_2S_{12}$) compounds. Currently, FIC are being investigated very intensively in the field of electrochemical devices since these materials are employed as electrolytes in solid-state batteries. The normal to superionic phase transition in FIC has associated a very large change of entropy (~100-200 $JK^{-1}kg^{-1}$) and can be modulated by external fields [90,91]. In particular, application of mechanical stresses has been proved as a very effective means to significantly lower or raise the superionic transition temperature in several archetypal FIC [92]. The physical causes behind these effects are the σ-induced lowering or increase in the kinetic energy barriers and formation energy of defects (e.g., Frenkel pairs) that intervene on ionic migration (Figure 5) [92]. As we explain next, such a mechanical control over the ionic conductivity, and simultaneously over the entropy, in superionic materials presents great prospects in the context of mechanocaloric solid-state cooling (Table 2).

FIC can be broadly classified into type-II and type-I categories, depending on whether the normal to superionic phase transition occurs in a continuous (type-II) or abrupt manner (type-I). $CaF_2$ is an archetypal type-II FIC with the cubic fluorite structure and a very high superionic transition temperature of ~ 1,350 K [89]. Giant EC effects of $|\Delta S|$ = 200 J $kg^{-1}$ $K^{-1}$ and $|\Delta T|$ = 163 K have been predicted in $CaF_2$ thin films under large tensile loads of ~ 5,000 MPa by means of atomistic-based simulation methods [70]. Despite that these theoretical findings in principle are of little practical relevance due to the high operation temperature involved, they come to show the great mechanocaloric potential that other similar materials with significantly lower superionic transition temperatures may have. This is the case, for instance, of $PbF_2$ and $Li_3N$, in which giant adiabatic temperature changes of 22 and 2 K have been predicted to occur at 600 K [70] and at room temperature [72], respectively. Likewise, giant room-temperature EC effects of $|\Delta S|$ = 30 J $kg^{-1}$ $K^{-1}$ and $|\Delta T|$ = 38 K have been estimated in AgI thin films (a prototype type-I FIC) under moderate compressive stresses of ~ 1,000 MPa [71]. In this latter case, the mechanocaloric effects are originated by a σ-induced diffusionless order-disorder phase transition [71,92].

A remarkable feature of type-II FIC is that possible refrigeration cycles based on them in principle should not suffer from the usual mechanical hysteresis issues affecting other families of EC materials (e.g., shape-memory alloys) since the normal to superionic phase transition is of second-order type and does not involve the nucleation of order-parameter domains or coexistence of structurally dissimilar phases. In addition, switching between the normal and superionic states can take place just within few picoseconds [93]. Consequently, refrigeration cycles based on type-II FIC could be performed at high repetition rates and with no appreciable degradation in cooling performance. On the other hand, mechanical fatigue issues are likely to appear due to the large changes in volume associated with the normal to superionic phase transition [89], which may facilitate the nucleation of and propagation of cracks and other defects.

## 4.4. Organometal halide perovskites

Organometal halide perovskites (OMHP) are at the frontier of renewable energy research because of their record speed of increasing photovoltaic efficiency and low economical costs associated to their synthesis [94]. The prototypical OMHP is methylammonium lead iodide with chemical formula $CH_3NH_3PbI_3$. OMHP present a perovskite structure $ABX_3$, analogous to that of many ferroelectric oxides, with an organic monovalent cation $MA^+$ at position A, a divalent metal at position B, and a halide anion at position X. The molecular cation has a permanent dipole and interacts with the inorganic $BX_3$ motifs through hydrogen bonding and van der Waals interactions [95]. A special feature of OMHP is that even at room temperature the A-site organic molecules can be orientationally disordered due to their small rotational energy barrier. Such molecular rotations preclude the appearance of long-range ferroelectric order [96], although switchable polar domains have been observed in a number of atomic force microscopy experiments [97]. It is likely then that OMPH, in analogy to ferroelectric relaxors, are spontaneously polarized at the nanoscale whereas polar-compensated at the macroscale. Interestingly, OMPH are good ionic conductors as well [98].

OMPH are mechanically soft as compared to oxide perovskites and can be stretched and compressed significantly even with moderate stresses. Recently, the room-temperature EC response of $CH_3NH_3PbI_3$ has been investigated in detail with simulation methods [73]. A giant adiabatic temperature change of 10.7 K has been predicted to occur for a moderate tensile uniaxial stress of 550 MPa (Table 2). The physical origins of such a giant EC effect is the σ-induced frustration of molecular rotations. Specifically, the $MA^+$ cations tend to align with the direction of the applied uniaxial stress hence reducing their rotation, which considerably reduces the entropy of the crystal. We should note, however, that the likely effects of ionic transport (i.e., iodine vacancies and interstitials diffusion) on the EC response of $CH_3NH_3PbI_3$ have been neglected in [73], and that in practice those might counteract the adiabatic temperature shift resulting exclusively from the $MA^+$ degrees of freedom (i.e., EC effects stemming from ionic transport changes are likely to be inverse for uniaxial tensile loads [70,72]). Direct or quasi-direct caloric measurements on OMPH, therefore, are highly desirable for unraveling their real EC potential. Nevertheless, as we explain later on, giant barocaloric effects have been already measured in OMPH near room temperature [99,100], which reassures their great caloric capability.

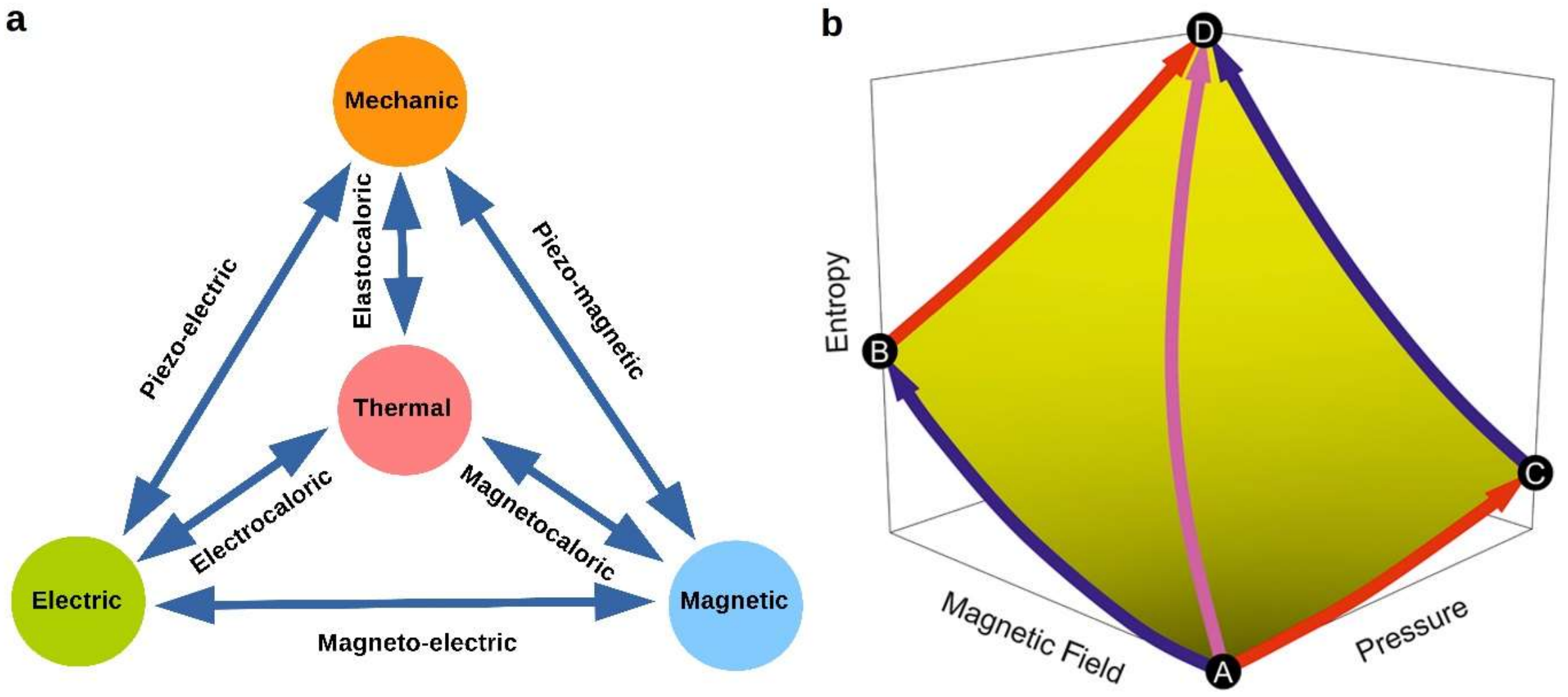

***Figure 6****: Multiferroic materials and multicaloric effects.* ***a.*** *Schematic diagram showing the possible mechanical, electrical and magnetic couplings in multiferroic materials. The couplings between different order parameters allows for multiple responses of the material upon application of a sole external field, thus producing multiple caloric effects.* ***b.*** *Schematic variation of isothermal entropy with pressure and magnetic field in a multiferroic material presenting coupling among lattice strain and magnetic order; multicaloric entropy changes can be obtained by applying pressure and magnetic field either sequentially or simultaneously (extracted from work [109]).*

## 4.5. Multiferroics

Multiferroics are materials in which several order parameters coexist, for example, ferroelectricity and magnetism, and can influence each other (that is, are coupled, Figure 6a). Multiferroics have been traditionally investigated in the context of electronic and logic devices and sensors and antennas [101-103], and as we explain next they also offer interesting possibilities as caloric materials [104-106]. First, in multiferroics different types of external fields can produce different caloric responses and this quality can be exploited to reduce noticeably detrimental hysteresis effects affecting solid-state refrigeration cycles [46]. And second, if the coupling between different order parameters is strong enough, one sole type of external stimulus can produce multiple caloric effects simultaneously, thus potentially enhancing the overall refrigerant performance. For these reasons, multiferroic materials can be also termed as "multicaloric". Several theoretical works have focused on establishing the thermodynamics of multicaloric effects [107-109] and in what follows we review their fundamentals.

Let us consider a multiferroic displaying *n* different types of ferroic orders $\{X_i\}$ ($i = 1,\ldots,n$) like polarization, magnetization, and lattice strain. For each ferroic property there is a ferroic field, $\{y_i\}$, that allows to change its value, like electric, magnetic field, and stress, and therefore make some work on the system. In multiferroics, the susceptibility tensor is defined as:

$$\chi_{ij} = \left(\frac{\partial X_i}{\partial y_j}\right)_{T,\{y_k \neq j\}}, \tag{14}$$

where the diagonal terms correspond to the standard susceptibilities and the cross-coefficients quantify the interplay between different ferroic orders. As we explain next, multicaloric effects occur when multiple fields are applied either sequentially or simultaneously on a multiferroic material, and in the case that the cross-coefficients $\chi_{ij}$'s are large also under the action of one sole field.

Let us examine a multiferroic in which the two ferroic orders $X_1$ and $X_2$ exists and are strongly coupled (i.e., $\chi_{12} \neq 0$). The isothermal entropy change, $\Delta S$, that results from applying the two external fields $y_1$ and $y_2$ on the multicaloric material, either simultaneously or sequentially, is [108,109]:

$$\Delta S\,[T,(0,0) \rightarrow (y_1,y_2)] = \Delta S\,[T,(0) \rightarrow (y_1)] + \Delta S\,[T,(0) \rightarrow (y_2)] + \int_0^{y_1}\int_0^{y_2} \left(\frac{\partial \chi_{12}}{\partial T}\right) dy_2 dy_1\,, \tag{15}$$

where the first two terms in the right-hand side correspond to usual caloric effects associated to thermodynamic conjugate variables $\{X_i, y_i\}$ by separate (i.e., mechanocaloric, electrocaloric, and magnetocaloric, depending on the nature of the applied external field and material). Likewise, the isothermal entropy that results from applying one sole type of external field, $y_2$, on the same multiferroic compound can be expressed as [108]:

$$\Delta S\,[T,(0,0) \rightarrow (0,y_2)\,] = \int_0^{y_2} \left[\left(\frac{\partial X_2}{\partial T}\right) + \frac{\chi_{12}}{\chi_{11}} \cdot \left(\frac{\partial X_1}{\partial T}\right)\right] dy_2\,. \tag{16}$$

To fix ideas, let us consider the explicit case of a mechanocaloric compound that is also ferroelectric (i.e., presents spontaneous and switchable electric polarization). In that particular case, the adiabatic temperature change that follows from applying a mechanical stress, according to Eq.(9) and (16), can be expressed analytically as:

$$\Delta T = -\frac{T}{C} \cdot \int_0^{\sigma} \left[ V_0 \left( \frac{\partial \varepsilon}{\partial T} \right) + \frac{d_e}{\chi_e} \left( \frac{\partial P}{\partial T} \right) \right] d\sigma \ . \qquad (17)$$

where $P$ represents the electric polarization, $d_e$ the piezoelectric coefficient ($\partial P/\partial \sigma$), and $\chi_e$ the electric susceptibility ($\partial P/\partial E$, where E represents electric field).

The third and second terms in Eqs.(15) and (16), respectively, offer the possibility of increasing in absolute value the isothermal entropy change, and therefore $\Delta T$, associated to the caloric responses. For instance, if the multicaloric material described by Eq.(17) is close to a ferroelectric-paraelectric phase transition, that is, when $\partial P/\partial T$ typically is largest, and possesses positive susceptibilities $d_e$ and $\chi_e$, which normally is the case, it is possible that the resulting $\Delta T$ is enhanced as compared to those found in analogous non-multiferroic materials. This is the case, for instance, of the giant multicaloric effects reported in the ferroelectric and antiferroelectric oxide perovskites $PbTiO_3$ [79] and $PbZrO_3$ [80], namely, $|\Delta T|$ ~ 35 and 25 K respectively (at quite high operating temperatures of 700 and 950 K), in which strong couplings between lattice strain and electric polarization exist. We should note, however, that in some cases the primary and secondary caloric effects, that is, those associated to the pairs of thermodynamic variables $\{X_i, y_i\}$ and $\{X_i, y_{j \neq i}\}$ respectively, may have opposite signs and therefore the overall caloric activity may turn out to be reduced [110].

Other facets in which multiferroic materials hold promise are related to the possibility of improving the degree of reversibility associated with caloric responses, and of exploiting thermal hysteresis, which in principle is adverse, in beneficial ways. As explained earlier in Section 3, the degree of reversibility accompanying a caloric process is determined by the hysteresis associated to the underlying phase transition, and the fraction of material that undergoes the transition both during cooling and subsequent heating in the cycle. Under the action of multiple external stimulus, the working temperature window associated with the caloric effect can be broaden so that the overlap between the succeeding cooling and heating processes is increased and reversibility is enhanced [111-112]. Meanwhile, hysteresis can be exploited in order to reduce considerably the magnitude of one type of external field in a multiple stimulus cycle. This is the case of $Ni_{49.6}Mn_{35.6}In_{14.8}$ Heusler alloys [46] and FeRh/BaTiO3 heterostructures [113] in which it has been shown that the intensity of the employed magnetic and electric fields, respectively, can be decreased significantly by combining them with mechanical stresses (Table 2). In the case of a ferroelectric compound presenting large piezoelectric coefficients, for instance, in principle it is possible to use compressive stress to make the polarization disappear (that is, to induce the ferroelectric-paraelectric phase transition) while using the electric field to recover the polarization by reverting the transition from the paraelectric to ferroelectric phase; in such a situation even modest electric fields may induce giant caloric effects [114].

# 5. BAROCALORIC MATERIALS

Barocaloric (BC) effects are driven by hydrostatic pressure, $P$. The analytical expression of the accompanying isothermal entropy changes, which is analogous to EC effects (Eq.8), is:

$$\Delta S = -\int_0^{P} \left( \frac{\partial V}{\partial T} \right)_{P'} dP', \qquad (18)$$

where $V$ represents the volume of the system. Large BC effects, therefore, typically occur near first-order transformations that render sizable volume changes. Next, we briefly review family of materials in which giant barocaloric effects (that is, $|\Delta T|$ ~ 10 K and $|\Delta T|$ ~ 100 J kg$^{-1}$ K$^{-1}$) have been discovered. We also highlight recent works in which unconventional barocaloric mechanisms, leading to significant caloric responses, have been reported.

| **Compound** | **Material type** | $T_t$ [K] | $\Delta P$ [MPa] | $\|\Delta S_t\|$ [J kg$^{-1}$ K$^{-1}$] | $\|\Delta T_{(qd)}\|$ [K] | $\|\Delta T\|/\Delta P$ [K MPa$^{-1}$] | **Reference** |
|---|---|---|---|---|---|---|---|
| $Ni_{51}Mn_{33}In_{16}$ | SMA | 330 | 250 | 41 | 4 | 0.02 | [115] |
| $Fe_{49}Rh_{51}$ | SMA | 310 | 110 | 12.5 | 9 | 0.08 | [116,117] |
| $BaTiO_3$ | FE | 400 | 100 | 2.4 | 1 | 0.01 | [118] |
| $(NH_4)_2SO_4$ | FE | 220 | 100 | 130 | 8 | 0.08 | [119] |
| [TPrA][Mn(dca)$_3$] | HP | 330 | 7 | 35 | 5 | 0.71 | [99,100] |
| AgI | FIC | 390 | 250 | 62 | 18 | 0.07 | [120] |
| $(CH_3)_2C(CH_2OH)_2$ | MC | 320 | 520 | 510 | 45 | 0.09 | [123,124] |
| [Fe(pzt)$_6$](PF$_6$)$_2$ | MC | 100 | 100 | 46 | 30 | 0.3 | [125] |
| [FeL$_2$][BF$_4$]$_2$ | MC | 262 | 30 | 80 | 3 | 0.1 | [126] |

***Table 3**: Materials in which giant BC effects have been measured. $T_t$ represents operating temperature, $\Delta P$ applied hydrostatic pressure, $\Delta S_t$ entropy change associated to the phase transition, and $\Delta T_{(qd)}$ adiabatic temperature change measured with quasi-direct methods. 'SMA', 'FE', 'HP', 'FIC', and 'MC' stand for shape-memory alloys, ferroelectrics, hybrid perovskites, fast-ion conductors, and molecular crystals, respectively. $|\Delta T|/\Delta P$ represents the barocaloric material strength, which is a good indicator of refrigerant performance.*

## 5.1. Overview of barocaloric performances

In Table 3, we enclose some of the most representative BC compounds along with some basic refrigerant characteristics. To date, giant BC effects have been measured in a number of organic-inorganic hybrid perovskites [99,100], shape-memory alloys [115-117], polar compounds [118,119], the archetypal fast-ion conductor AgI [120], fluoride-based materials [121], polymers [122], and molecular crystals [123-126]. The phase transitions leading to the giant BC effects reported in Table 3 present some similarities to those explained in preceding sections, made the exception of molecular crystals. For instance, for a small hydrostatic pressure (~100 MPa) the two shape memory alloys $Ni_{51}Mn_{33}In_{16}$ and $Fe_{49}Rh_{51}$ undergo magneto-structural transformations, the two ferroelectric materials $BaTiO_3$ and $(NH_4)_2SO_4$ polar to non-polar phase transitions, and the rest of compounds purely structural transformations. Arguably, the less good BC performer is the oxide perovskite $BaTiO_3$ (e.g., low barocaloric strength $|\Delta T|/\Delta P$ and high operating temperature) which just experiences a small relative change in volume of 0.11% during the tetragonal to cubic phase transition [118] (see Sec. 4.2). On the other hand, quite remarkable results are the barocaloric strength and giant adiabatic temperature change measured in the hybrid perovskite [TPrA][Mn(dca)$_3$] and fast-ion conductor AgI, respectively (Table 3). In the latter system, a record refrigerant cooling performance (*RCP*, see Sec. 2.1) of 2.5 kJ $kg^{-1}$ has been measured for $P$ = 250 MPa, which results from the existence of large BC effects over an ample temperature span of ~ 60 K [120].

The adiabatic temperature changes measured with quasi-direct methods in BC materials are slightly smaller, although comparable in magnitude, to those measured with direct methods in EC materials (i.e., quantities $|\Delta T_{(qd)}|$ and $|\Delta T_{(d)}|$ in Tables 3 and 1, respectively). We note that when comparing different caloric effects in different materials one needs to pay special attention on how the experimental data have been determined. For instance, the adiabatic temperature changes that are estimated straightforwardly from the entropy changes associated to the phase transitions normally are much higher than those measured with direct or quasi-direct methods (e.g., compare the $|\Delta T_t|$ and $|\Delta T_{(q)}|$ values in Table 1). In BC materials, the discrepancies between $|\Delta T_t|$ and $|\Delta T_{(qd)}|$ can be especially large (e.g., they amount to ~ 30 K in $Ni_{51}Mn_{33}In_{16}$ [10]) due to two main causes. First, the lack of adiabaticity associated with the pressure transmitting medium which may not be perfectly isotropic. And second, the partial completion of the involved phase transition owing to phase coexistence and hysteresis effects typically accompanying first-order transformations. Such adverse effects, which may lead to limiting irreversibility issues, in principle are not so critical in EC materials since pressure transmitting media are not required and the involved phase transitions can be driven to completion more efficiently. An illustrative example of this aspect is the EC effect observed in Ni-Ti alloys, which is one of the largest reported to date and for which the differences between $|\Delta T_t|$ and $|\Delta T_{(d)}|$ are practically negligible (Table 1) [10].

## 5.2. Recent developments

The barocaloric strengths and giant adiabatic temperature changes measured in molecular crystals, reported within the last few years, are very impressive. For instance, the $|\Delta T_{(qd)}|$ obtained in neopentylglycol, with chemical formula $(CH_3)_2C(CH_2OH)_2$ [1,123,124], amounts to 45-50 K near room temperature, which actually deserves the adjective "colossal". On the other hand, the $\Delta T/\Delta P$ values reported in the magnetic compounds [Fe(pzt)$_6$](PF$_6$)$_2$ [125] and [FeL$_2$][BF$_4$]$_2$ [126] are superior to those achieved in the rest materials listed in Table 3, made the exception of the hybrid perovskites [TPrA][Mn(dca)$_3$] (although their corresponding operating temperatures probably are too low). Next, we explain in more detail the atomistic mechanisms leading to such large and promising caloric effects in molecular crystals.

Plastic crystals are molecular solids in which the interactions between molecules are very weak and long-ranged (that is, dispersion like). As a consequence, plastic crystals are highly compressible and can be deformed in a reversible manner, hence the adjective "plastic". Under certain pressure and temperature conditions, molecules in plastic crystals can start rotating almost freely around their centres of mass. The centres of mass remain localized at well-defined and ordered positions in the crystal lattice while molecular rotation leads to orientational disorder and a high-entropy state. By means of external pressure it is possible to stop such molecules rotations, hence inducing a fully ordered state with low entropy. Very recently, it has been shown by several independent research groups [123,124] that the entropy changes driven by pressure near 300 K in plastic crystals are huge and render colossal BC effects (see $(CH_3)_2C(CH_2OH)_2$ in Table 3). It is worth mentioning that similar $\sigma$-induced caloric mechanisms were predicted previously in $CH_3NH_3PbI_3$ based on the outcomes of classical molecular dynamics simulations [73] (Section 4.4). In view of the high anharmonicity and mechanical softness of plastic crystals, and of the results reported previously in organometal halide perovskites [73], it is very likely that analogous caloric responses induced by uniaxial stresses, leading to huge EC effects, could be achieved. Figure 7 sketches a possible 4-steps cooling cycle based on the colossal BC effects disclosed in plastic crystals.

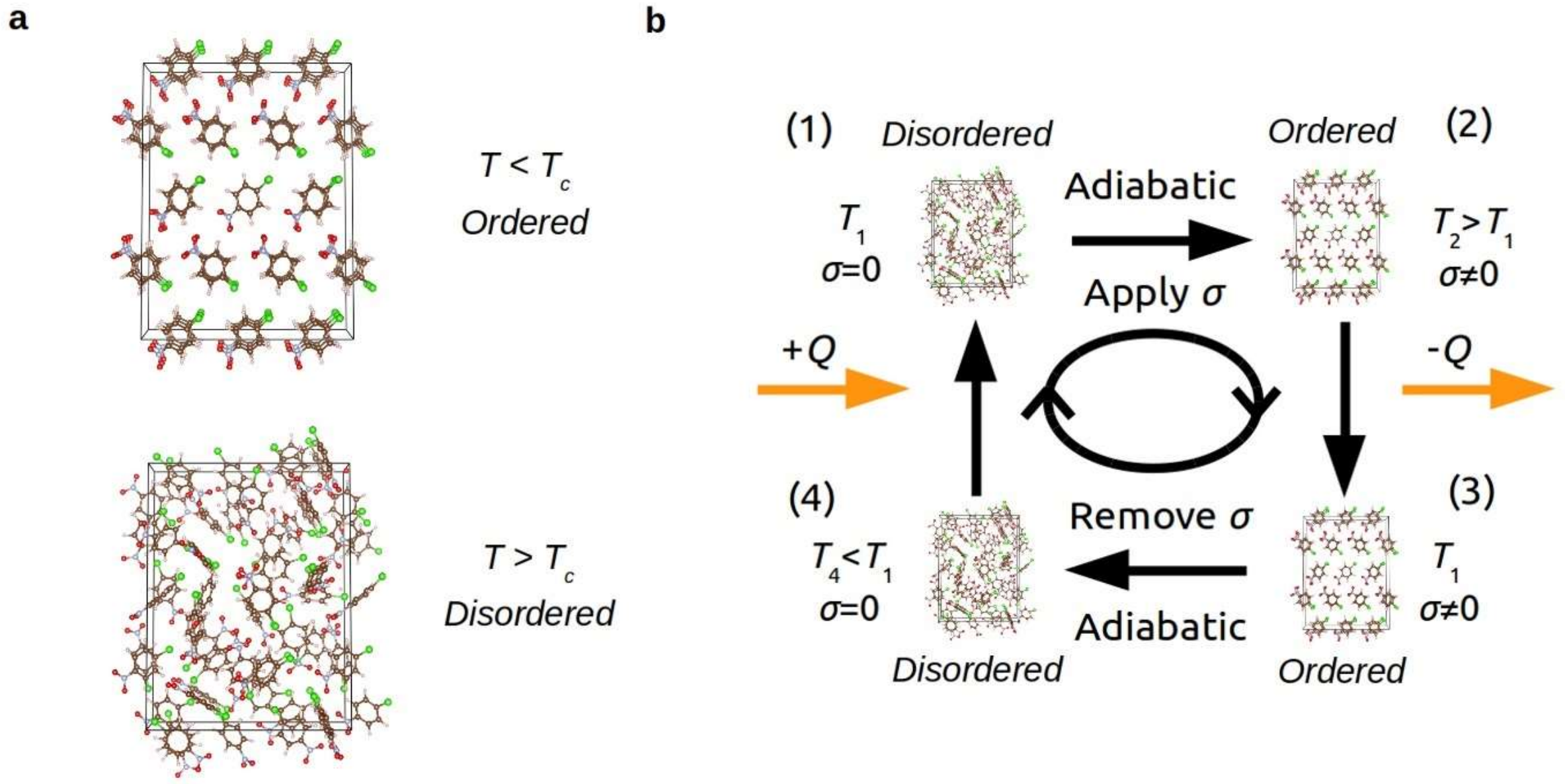

***Figure 7****: The molecular order-disorder phase transition in plastic crystals and their potential use in solid-state cooling.* ***a****. The molecular ordered and disordered phases of a plastic crystal are sketched; the disordered high-entropy phase is stabilised at high temperatures while the ordered low-entropy phase is stabilised at low temperatures.* ***b****. Proposal of a simple four-step refrigeration cycle based on external pressure,* $\boldsymbol{\sigma}$*, and plastic crystals; (1) initially the crystal is in the molecular disordered phase; (2) pressure is applied adiabatically on the crystal so that the ordered phase is stabilised and the temperature of the crystal increases; (3) heat is ejected from the system and the temperature of the crystal returns to its initial value; (4) pressure is released adiabatically from the crystal so that the disordered phase is recovered and the temperature of the crystal decreases; the cycle is completed by absorbing heat from a hot source and returning to the initial temperature.*

Plastic crystals are quite different from other caloric materials, and not just because of their huge entropy changes near room temperature: they are cheap and easy to produce, lightweight, non-toxic, and flexible. Plastic crystals, therefore, seem to be especially well suited for the integration of solid-state cooling in electronic devices and mobile applications. Nevertheless, plastic crystals also suffer from some important problems in the context caloric materials. For instance, given their organic nature, plastic crystals have relatively low melting points (typically about 300-400 K [127]), which is not desirable for refrigeration applications. In addition, the same properties that make plastic crystals highly deformable imply that these materials also lack the mechanical resilience to endure many refrigeration cycles. Perhaps most importantly, hysteresis and phase-coexistence effects are likely to weaken the cooling performance of plastic crystals due to the first-order nature of the molecular order-disorder phase transformation. A possible solution to address this latter technical issue may consist in combining multiple external stimulus (see Section 4.5), for instance, electric fields and mechanical stresses.

Recently also, large BC effects have been predicted and subsequently demonstrated experimentally in magnetic materials undergoing spin-crossover (SCO) phase transitions. Spin-crossover phase transitions refer to transformations from a low-spin state to a high-spin state as induced by increasing temperature, which typically are accompanied by large volume changes and may occur near or beyond room temperature [128]. By applying and removing hydrostatic pressure on SCO compounds (typically ~0.1-0.01 GPa) it is also possible to trigger the transition from low-spin to high-spin state and *vice versa*. Archetypal SCO materials are molecular crystals that contain magnetic transition metal ions (e.g., Fe) [125,126], and the usual order of magnitude of the entropy changes associated with their transformations is ~100 J $kg^{-1}$ $K^{-1}$ [128]. The origins of such large entropy changes are linked to both magnetic spin ordering and lattice vibrations, which are very likely to be strongly coupled among them. However, a microscopic design approach able to optimize those entropy increments rationally has not emerged yet, presumably due to the lack of related theoretical studies based on first-principles methods (in which the magnetic spin and lattice degrees of freedom in principle can be described accurately and on an equal footing [39,105]). A potential design benefit of SCO compounds is that they are highly tuneable in the sense that by changing some of their molecular building motifs the critical temperature and volume change accompanying the spin-state transition can be modified appreciably [128,129].

## 6. CONCLUDING REMARKS AND OUTLOOK

Solid-state cooling is an environmentally friendly, energy-efficient and highly scalable technology that may solve most of the problems associated with current refrigeration methods based on gas compression-decompression cycles (including, potential toxicity and green-house effects). It relies on applying external magnetic, electric, mechanical fields, or a combination of them, on compounds that undergo temperature variations as a result of field-induced phase transitions that involve large changes in entropy. Mechanocaloric effects, either produced by uniaxial stress or hydrostatic pressure, render the largest adiabatic temperature variations of all caloric responses, typically $\Delta T \sim 10$ K, hence are particularly promising from an applied perspective. In this regard, elastocaloric effects probably are the most encouraging since do not require pressure-transmitting media, the involved phase transitions can be driven to completion more easily, and both compressive and tensile stress can be explored thus *a priori* richer caloric responses can be attained (in contrast to hydrostatic pressure which in practice can only be compressive). Furthermore, most mechanocaloric compounds do not suffer from limiting issues caused by the application of external fields (e.g., leakage currents and dielectric losses in electrocaloric materials) and are relatively abundant in nature (in contrast to prototype magnetocaloric materials which contain rare-earth elements).

In the present review, we have explained the key properties of archetypal and non-conventional families of mechanocaloric materials (e.g., NiTi-based shape memory alloys and superionic compounds, respectively). Although conventional mechanocaloric materials already display superb cooling performances (e.g., $\Delta T = 25$ K near room temperature upon application-removal of ~500 MPa uniaxial tensile stress in Ni-Ti wires [43,44]) they are not perfect and present also few important drawbacks. Such limitations mainly are related to hysteresis and materials fatigue issues, leading to irreversibility and poor cyclability, which may compromise seriously the performance in refrigeration cycles. Nevertheless, a number of new mechanocaloric materials and ingenious multiple stimulus strategies have been proposed recently that can potentially overcome those problems and thus advance the field of solid-state cooling.

On the materials side, it is possible to find mechanocaloric compounds exhibiting second-order, or weak first-order, phase transitions which by definition entail small hysteresis and can be triggered somehow reversibly. Examples of such materials include inorganic ferroelectrics (e.g., oxide perovskites) and type-II fast-ion conductors (e.g., $PbF_2$ and $Li_3N$). In the particular case of type-II fast-ion conductors, mechanocaloric effects are originated by order-disorder phase transitions affecting the sublattice of mobile ions as the material enters the superionic state. Nevertheless, the operating temperatures associated with some ferroelectrics and fast-ion conductors normally are well above or below room temperature, which suggests the use of additional doping strategies in order to shift the corresponding transition temperatures. Meanwhile, colossal mechanocaloric effects ($\Delta T$ ~ 50 K near room temperature for small hydrostatic pressures of ~ 0.1 GPa [123,124]) have been discovered very recently in a class of molecular solids known as plastic crystals (e.g., neopentylglycol), which are also related to pressure-driven order-disorder phase transitions (it remains to be demonstrated whether analogous elastocaloric effects exist also in such materials). Interestingly, other types of molecular solids like organic-inorganic hybrid perovskites and spin-crossover complexes have been reported to exhibit also giant mechanocaloric effects in the last few years. In view of the very encouraging mechanocaloric effects observed in molecular and disordered crystals and of their fresh presence in the context of caloric effects, there seems to be a lot of room for possible advancements based on them in the near future.

On the multiple external stimulus side, it has been shown also recently that the combination of multiple external fields, applied either sequentially or simultaneously, on multiferroic materials (typically, magnetic Heusler alloys and oxide perovskites) may be very advantageous in terms of enhanced reversibility, enhanced caloric response, and reduction in the intensity of coercive fields (this last point turns out to be especially relevant for magnetocaloric materials [46,113]). For these multiple external field approaches to succeed, it is crucial to find suitable multiferroic materials in which several order parameters coexist and (ideally) are strongly coupled. In customary multiferroic compounds, lattice strain appears most of the times intertwined with the polar or magnetic degrees of freedom hence one of the multiple external fields should preferably be mechanical stress (leading to combined electric field-mechanical stress or magnetic field-mechanical stress cycles). Research on multicaloric effects and multicaloric cycles is in its early stage and there seems to be also a lot of promise along that refrigeration materials direction.

The present review on non-conventional mechanocaloric materials and mechanical effects aims to motivate new theoretical and experimental research on solid-state cooling. Our critical assessment of materials suggests that mechanocaloric technologies could benefit immensely from the intensive research already undertaken in the fields of functional materials (multiferroic, ferroelectric, and magnetic compounds), energy storage (fast-ion conductors), and coordination chemistry (plastic and molecular crystals).